\documentclass{svproc}
\usepackage{url}
\usepackage{graphicx}
\usepackage[labelfont={small},subrefformat=parens,caption=false]{subfig}

\begin{document}
\mainmatter              

\newcommand{\mc}{M_{\mathrm{3}}}
\newcommand{\mn}{M_{\mathrm{4}}}
\newcommand{\mnc}{M_{\mathrm{34}}} 
\newcommand{\mr}{M_{\mathrm{R}}}
\newcommand{\gma}{\gamma}
\newcommand{\gmast}{\gamma_\ast}
\newcommand{\pc}{\vec{p}_{c}}
\newcommand{\pn}{\vec{p}_{n}}
\newcommand{\bes}{{}^{7}\mathrm{Be}}
\newcommand{\besst}{{}^{7}\mathrm{Be}^\ast}
\newcommand{\be}{{}^{8}\mathrm{B}}
\newcommand{\het}{{}^{3}\mathrm{He}}
\newcommand{\hef}{{}^{4}\mathrm{He}}
\renewcommand{\S}[2]{{}^{}S_{#2}}
\renewcommand{\P}[2]{{}^{}P_{#2}}
\newcommand{\gone}{g_{(\S{2}{1/2})}}
\newcommand{\aone}{a_{(\S{2}{ 1/2})}}
\newcommand{\hone}{h_{(\P{2}{3/2})}}
\newcommand{\htwo}{h_{(\P{2}{1/2})}}
\newcommand{\V}[1]{\vec{V}_{#1}}
\newcommand{\fdu}[2]{{#1}^{\dagger #2}}
\newcommand{\fdd}[2]{{#1}^{\dagger}_{#2}}
\newcommand{\fu}[2]{{#1}^{#2}}
\newcommand{\fd}[2]{{#1}_{#2}}
\newcommand{\T}[2]{T_{#1}^{\, #2}}
\newcommand{\e}{\vec{\epsilon}}
\newcommand{\es}{\e^{*}}
\newcommand{\cw}[2]{\chi^{(#2)}_{#1}}
\newcommand{\cwc}[2]{\chi^{(#2)*}_{#1}}
\newcommand{\cwf}[1]{F_{#1}}
\newcommand{\cwg}[1]{G_{#1}}
\newcommand{\ke}{k}
\newcommand{\kest}{k_{\ast}}
\newcommand{\kc}{k_{C}}
\newcommand{\upartial}[1]{\partial^{#1}}
\newcommand{\dpartial}[1]{\partial_{#1}}
\newcommand{\etae}{\eta }
\newcommand{\etab}{\eta_{B}}
\newcommand{\etaest}{\eta_{ \ast}}
\newcommand{\etabst}{\eta_{B\ast}}
\newcommand{\vecpt}[1]{\hat{\vec{#1}}}
\newcommand{\uY}[2]{Y_{#1}^{#2}}
\newcommand{\dY}[2]{Y_{#1 #2}}

\def\lsim{\mathrel{\rlap{\lower4pt\hbox{\hskip1pt$\sim$}}
    \raise1pt\hbox{$<$}}}         
\def\gsim{\mathrel {\rlap{\lower4pt\hbox{\hskip1pt$\sim$}}
    \raise1pt\hbox{$>$}}}         

\title{$S$-factor and scattering parameters from\\ ${}^3$He + ${}^4$He $\rightarrow {}^7$Be + $\gamma$ data}
\titlerunning{$S$-factor from ${}^3$He + ${}^4$He $\rightarrow {}^7$Be + $\gamma$}  
%
\author{Xilin Zhang \inst{1,2} \and Kenneth M. Nollett \inst{3} \and Daniel R. Phillips \inst{4,5,6}}
\authorrunning{Xilin Zhang et al.} 
%
\tocauthor{Xilin Zhang and Kenneth M. Nollett and Daniel R. Phillips}
\institute{Physics Department, University of Washington, 
Seattle, WA 98195, USA, \\ 
\and
Department of Physics, The Ohio State University, Columbus, OH 43210, USA, \\ \email{zhang.10038@osu.edu}
\and
Department of Physics, San Diego State University,
5500 Campanile Drive, San Diego, California 92182-1233, USA,\\ \email{kenneth.nollett@sdsu.edu}
\and
Institute of Nuclear and Particle Physics and Department of
Physics and Astronomy, Ohio University, Athens, OH\ \ 45701, USA,\\ 
\and
Institut f\"ur Kernphysik, TU Darmstadt, 64289 Darmstadt, Germany,
\and
ExtreMe Matter Institute EMMI, GSI Helmholtzzentrum f{\"u}r Schwerionenforschung GmbH, 64291 Darmstadt, Germany \\ \email{phillid1@ohio.edu}
}
\maketitle              

\begin{abstract}
 We use the next-to-leading-order (NLO) amplitude in an effective field theory (EFT) for $\het +\hef \rightarrow \bes + \gamma$ to perform the extrapolation of higher-energy data to solar energies. At this order the 
EFT describes the capture process
using an s-wave scattering length and effective range, the asymptotic behavior of $\bes$ and its excited state, and short-distance contributions to the E1 capture amplitude. We use a Bayesian analysis
to infer the multi-dimensional posterior of these parameters from capture data below 2 MeV. The total $S$ factor $S(0)= 0.578^{+0.015}_{-0.016}$ keV b
at 68\% degree of belief. We also find significant constraints on $\het$-$\hef$ scattering parameters. 
\end{abstract}

\noindent
The solar-fusion reaction $\het +\hef \rightarrow \bes + \gamma$ has not been measured directly at solar energies, due to the exponential suppression of the cross section there.  Solar models use cross sections for it  based on extrapolants that are derived using potential models or R-matrix, and constrained by 
$S$-factor and $\het$-$\hef$ scattering data, as well as $\bes$ bound-state properties.  Ref.~\cite{Adelberger:2010qa} reviews the most prominent efforts before
2011; additional evaluations have emerged since~\cite{deBoer:2014hha,Iliadis:2016vkw}.

\section{Formalism for $E1$ capture} 
\label{sec:E1form}

We use Halo Effective Field Theory (EFT)~\cite{Hammer:2017}, treating $\het$ ($\frac{1}{2}^+$) and $\hef$ ($0^+$) as fundamental degrees of freedom and $\bes$ (ground state, GS, $\frac{3}{2}^-$) and $\bes^\ast$ (excited state, ES, $\frac{1}{2}^-$) as shallow p-wave  bound states of the two. From the breakup energies of $\het$ and $\hef$ we infer an EFT breakdown scale $\Lambda$ of about $200$ MeV. The energy range $E \lsim 2$ MeV implies a low-momentum scale $Q$ of $70$--$80$ MeV, thus we have $Q/\Lambda \approx 0.4$. We systematically expand both scattering and reaction amplitudes in this small parameter.
The NLO $S$-factor for E1 capture to the ${}^7$Be GS can then be written \cite{Zhang:2019}
\begin{equation}
S_{_{\P{2}{3/2}}}(E) =  \frac{e^{2\pi \etae}}{e^{2\pi\etae}-1}  \frac{8\pi}{9}  \left(e\, Z_{eff}\right)^{2}  \kc \omega^3 C_{(\P{2}{3/2})}^{2} \left(\mid \mathcal{S} \mid^{2}+2 \mid \mathcal{D} \mid^{2}\right) \ , \label{eqn:sfactormaster1} 
\end{equation}
with the same result, {\it mutatis mutandis} for capture to the $\P{2}{1/2}$ ES. This is analogous to our results for $\bes + p \rightarrow \be + \gma$~\cite{Zhang:2015ajn,Zhang:2017yqc}. 
Here, $\kc\equiv \alpha_\mathrm{em} Z^2 \mr$ with $\mr$  the reduced mass of the $\het$-$\hef$ system; $\eta\equiv \kc/p$ is the well-known Sommerfeld parameter; $\omega$ is the energy of the photon produced in the
reaction; and the ``effective charge" $Z_{eff} \equiv \left(Z/\mn-Z/\mc\right) \mr$.
The factors $C_{(\P{2}{3/2})}^{2}$  ($C_{(\P{2}{1/2})}^{2}$) are the squared p-wave asymptotic normalization coefficients (ANCs) of the GS (ES)~\cite{Zhang:2017yqc}.
The two reduced matrix elements, $\mathcal{S}$ and $\mathcal{D}$, are for the E1 transition from initial s- and d-wave states. At NLO, $\mathcal{S}$ consists of the well-known external capture contributions plus a short-distance piece similar to R-matrix internal capture. We parameterize the latter contribution to capture to the GS (ES) by a single number, $\overline{L}$ ($\overline{L}_\ast$). 
The d-wave reduced matrix element $\mathcal{D}$ is given by the standard asymptotic expression for external capture,  but integrated all the way to zero radius. Explicit formulae for $\mathcal{S}$ and $\mathcal{D}$ can be found in Refs.~\cite{Zhang:2019,Zhang:2017yqc}.

%
Capture reactions to the ground and excited state share the same initial state for s-waves ($\frac{1}{2}^+$), so $\mathcal{S}$ depends on the scattering length, $a_0$ and effective range, $r_0$. 
Up to NLO there are then 6 EFT parameters, henceforth denoted as the vector ${\bf g}$: $ C_{(\P{2}{3/2})}^2$ ($\mathrm{fm}^{-1}$), $ C_{(\P{2}{1/2})}^2$ ($\mathrm{fm}^{-1}$), $a_0$ (fm), $r_0$ (fm), $\overline{L}$ (fm), and $\overline{L}_\ast$ (fm).  

\section{Data, Bayesian analysis, and Results}

\begin{figure}[htb]
\centering
\includegraphics[width=0.65\textwidth]{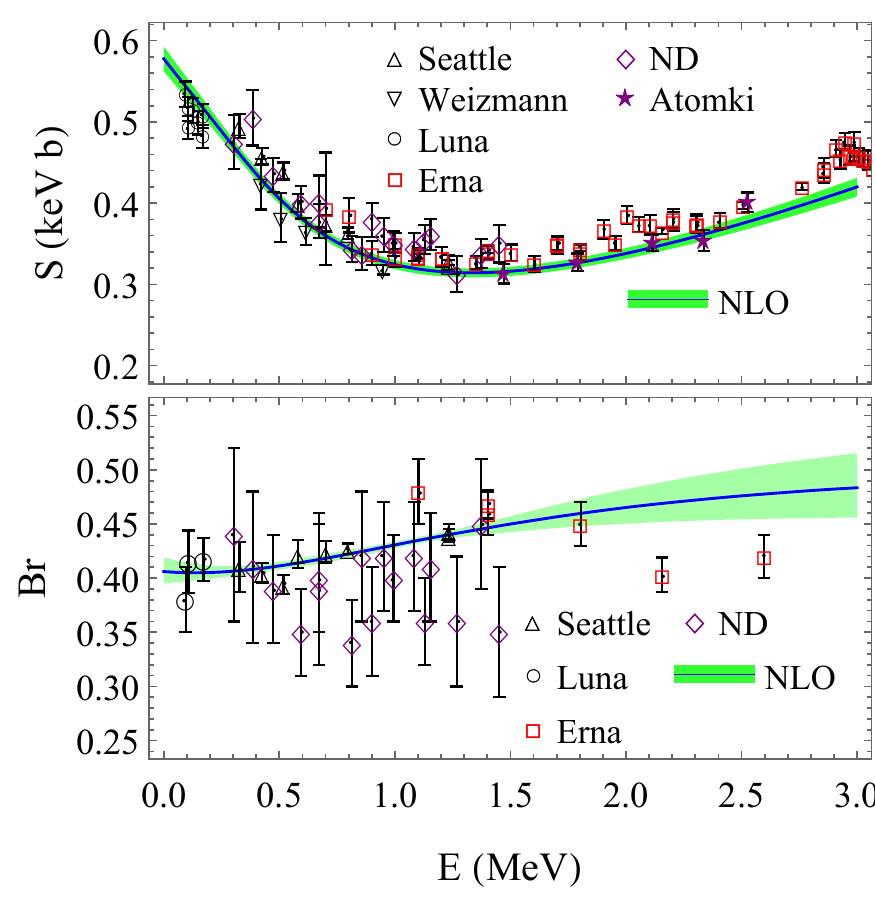}
\vspace{-0.4cm}
\caption{Total $S$-factor and branching-ratio results. The data is denoted  in the legend, and summarized in Ref.~\cite{deBoer:2014hha}. The green band shows the 68\% interval for $S(E)$ and $Br(E)$ in our NLO Halo EFT analysis. The mean is denoted by the blue  line.} \label{fig:SBrvsENLON4Lv2FullData}
\end{figure}

There are six total $S$-factor data sets, here labeled Seattle, Weizmann, Luna, Erna, Notre Dame (ND), and Atomki.  There are four branching-ratio data sets: Seattle, Luna, Erna, and Notre Dame.  In order to ensure that the data used are within the domain of validity of the EFT we only employ data with $E \leq 2$ MeV. This, together with other data-selection criteria, yields 59 $S$-factor and 32 $Br$ data, see Fig.~\ref{fig:SBrvsENLON4Lv2FullData}. (Details, including original references and a full listing of these data, will appear in Ref.~\cite{Zhang:2019}.)  To account for the common-mode errors we introduce normalization corrections, \{$\xi_J$: $J=1 \ldots N_\mathrm{exp}$\}, for the $S$-factor data. Such errors mostly cancel for $Br$ data, so this correction is not used for them.

We take the EFT expressions such as (\ref{eqn:sfactormaster1}) 
and employ Bayesian analysis---implemented via Markov-Chain-Monte-Carlo (MCMC) sampling---to infer probability distribution functions (PDFs) for the EFT parameters ${\bf g}$ from these data. Taking box priors with ranges considerably larger than those suggested by naive dimensional analysis for ${\bf g}$, and gaussian priors for the $\xi_J$'s, we can write the desired PDF as $ {\rm pr} \left({\bf g},\{\xi_J\} \vert D;T; I \right) \equiv c\, \exp\left(-\chi^2/2\right) $.
The $\chi^2$ is non-standard because it includes not only contributions from $S$-factor and branching-ratio measurements, but also the effect of the normalization corrections~\cite{Zhang:2019,Zhang:2017yqc}.

\vspace{-0.5cm}

The MCMC sampling produces the full twelve-dimensional pdf for $\vec{g}$ and $\{\xi_J\}$. These samples can then be used to compute a histogram for $S(E)$ and $Br(E)$ at any energy, $E$.
Fig.~\ref{fig:SBrvsENLON4Lv2FullData} shows the resulting 68\% intervals: the mean is denoted by the blue line. The data is shown without re-scaling by the factors $1/(1-\xi_J)$, so Fig.~\ref{fig:SBrvsENLON4Lv2FullData} under-represents the quality of our final result. 
Adopting values for the $\xi_J$'s that maximize their posterior pdf produces a distribution of $\chi^2$'s in our MCMC sample peaks at 1.1 per degree of freedom.

At NLO we have 
$S(0)= 0.578^{+0.015}_{-0.016}$ keV b. The recommended value  from Ref.~\cite{Adelberger:2010qa} is $0.56\pm 0.02\mathrm{(exp)} \pm 0.02\mathrm{(theory)}$---consistent with our result, but with an uncertainty that is almost a factor of two larger. Other recent analyses are broadly consistent, but also have somewhat bigger errors~\cite{deBoer:2014hha,Iliadis:2016vkw}. We also find
$Br(0)=0.406^{+0.013}_{-0.011}$. There are two essential differences between this paper and another, recent, EFT evaluation of the same reaction~\cite{Higa:2016igc}.  First, we employ Bayesian methods. Second, we do not not include existing scattering phase shift analyses in our constraints because their systematic errors are poorly quantified.

\begin{figure}
\centering
\includegraphics[width=0.39 \textwidth]{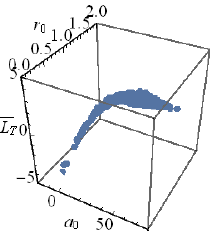}
\vspace{-0.4cm}
\caption{$a_0$--$r_0$--$\overline{L}_T$ (all in fm) 3D scatter plot based on MCMC samples.} \label{fig:a_r_Lt_Corr_NLOFullData}
\end{figure}

Fig.~\ref{fig:a_r_Lt_Corr_NLOFullData} displays a three-dimensional scatter plot of the NLO MCMC samples, projected to the $a_0$--$r_0$--$\overline{L}_T$ subspace. Projecting further onto the $a_0$-$r_0$ subspace shows that significant constraints on $\het$-$\hef$ scattering parameters can be obtained from the extant radiative capture data---in contrast to cases such as $\bes(p,\gamma)$~\cite{Zhang:2015ajn}. The corresponding effective-range function can be tested against future high-quality $\het$-$\hef$ scattering data at low energy.

We conclude that data on $\het +\hef \rightarrow \bes + \gamma$ already tightly constrain important aspects of the dynamics needed for extrapolation of this reaction's $S$-factor: we find quite small uncertainties on the s-wave elastic scattering parameters and the ANCs of the final states.  Better measurements of scattering cross sections will test the EFT approach to the reaction presented here.

\paragraph{Acknowledgments} This work is supported by the US Department of Energy under grant no. DE-FG02-93ER-40756 (DP), DE-FG02-97ER-41014 (XZ), DE-SC0019257 (KMN), and through MSU subcontract RC107839-OSU for the NUCLEI SciDAC collaboration (XZ), by the US National Science Foundation under Grant No. PHY-1614460 (XZ), and by the 
 ExtreMe Matter Institute EMMI at the GSI Helmholtzzentrum f\"ur Schwerionenphysik (DP). 
%
%

\end{document}